\documentclass[fleqn, usenatbib]{mnras}

\usepackage{amsmath}
\usepackage{physics}

\newcommand{\lya}{Ly$\alpha$}

\DeclareRobustCommand{\VAN}[3]{#2}
\let\VANthebibliography\thebibliography
\def\thebibliography{\DeclareRobustCommand{\VAN}[3]{##3}\VANthebibliography}

\usepackage{amssymb}
\usepackage{graphicx}

\title[Predicting \lya{} Emission with Neural Network]{Predicting \lya{} Emission from Distant Galaxies with Neural Network Architecture}

\author[T. Yoshioka et al.]{
    Takehiro Yoshioka,$^{1}$\thanks{E-mail: yoshioka@astron.s.u-tokyo.ac.jp}
    Nobunari Kashikawa,$^{1,2}$
    Yoshihiro Takeda,$^{1}$
    Kei Ito,$^{1,3,4}$ \newauthor
    Yongming Liang,$^{5}$
    Rikako Ishimoto,$^{1}$
    Junya Arita,$^{1}$
    Yuri Nishimura,$^{6}$ \newauthor
    Hiroki Hoshi,$^{1}$
    Shunta Shimizu,$^{1}$
    \\
    $^{1}$Department of Astronomy, School of Science, The University of Tokyo, 7-3-1 Hongo, Bunkyo-ku, Tokyo, 113-0033, Japan\\
    $^{2}$Research Center for the Early Universe, The University of Tokyo, 7-3-1 Hongo, Bunkyo-ku, Tokyo, 113-0033, Japan\\
    $^{3}$Cosmic Dawn Center (DAWN), Denmark\\
    $^{4}$DTU Space, Technical University of Denmark, Elektrovej 327, DK2800 Kgs. Lyngby, Denmark\\
    $^{5}$Institute for Cosmic Ray Research, The University of Tokyo, Kashiwa, Chiba 277-8582, Japan\\
    $^{6}$Institute of Astronomy, Graduate School of Science, The University of Tokyo, 2-21-1 Osawa, Mitaka, Tokyo 181-0015, Japan
}

\date{Accepted 2024 December 11. Received 2024 December 11; in original form 2024 August 10}

\pubyear{2024}

\begin{document}
\label{firstpage}
\pagerange{\pageref{firstpage}--\pageref{lastpage}}
\maketitle

\begin{abstract}
    The \lya{} emission line is a characteristic feature found in high-$z$ galaxies, serving as a probe of cosmic reionization.
    While previous works present various correlations between \lya{} emission and physical properties of host galaxies, it is still unclear which characteristics predominantly determine the \lya{} emission.
    In this study, we introduce a neural network approach to simultaneously handle multiple properties of galaxies.
    The neural-network-based prediction model that identifies \lya{} emitters (LAEs) from six physical properties: star formation rate (SFR), stellar mass, UV absolute magnitude $M_\mathrm{UV}$, age, UV slope $\beta$, and dust attenuation $E(B-V)$, obtained by the SED fitting. 
    The network is trained with galaxy samples from the VANDELS and MUSE spectroscopic surveys and achieves the performance of 77\% true positive rate and 14\% false positive rate.
    The permutation feature importance method shows that $\beta$, $M_\mathrm{UV}$, and $M_*$ are important for the prediction of LAEs.
    As an independent validation, we find that 91\% of LAEs spectroscopically confirmed by the James Webb Space Telescope (JWST) have a probability of LAE higher than 70\% in this model.
    This prediction model enables the efficient construction of a large LAE sample in a wide and continuous redshift space using only photometric data.
    We apply the prediction model to the JWST photometric galaxy sample and obtain \lya{} fraction consistent with previous studies.
    Moreover, we demonstrate that the difference between the distributions of LAEs predicted by the model and the spectroscopically identified LAEs provides a strong constraint on the H\,\textsc{ii} bubble size.
\end{abstract}


\section{Introduction}\label{sec:introduction}

\lya{} emission, theoretically predicted by \citet{Partridge67}, is one of the powerful probes of high-$z$ Universe.
Significant efforts are invested in detecting \lya{} emission because of its ease of observability as one of the strongest intrinsic features in high-z galaxy spectra.
\lya{} is also a powerful tool for exploring cosmic reionization, given its nature of being scattered by neutral hydrogen gas.
\lya{} fraction, defined as the number ratio of LAEs to total galaxies, is one of the frequently used methods for investigating cosmic reionization.
Previous studies have reported that \lya{} fraction sharply drops from $z=6$ to $z=7$, while the continuous increase at $4 < z < 6$ \citep{Stark11, Schenker14}.
This is thought to be the indicator of the neutral IGM in the epoch of reionization (EoR).
Recent work continues to derive \lya{} fraction \citep{Jones23, Goovaerts23, Bolan22, Fuller20}, constraining the evolution history of the neutral fraction.

Increasing the sample size of \lya{} emitters (LAEs) is important for carrying out a wide range of studies in high-$z$ Universe.
However, spectroscopic observations, while a valuable tool for identifying \lya{}, require a long observing time, making it challenging to increase the sample size.
Although the LAE selection using narrow-band (NB) filter observation allows for efficient surveys over a wide field of view, the wavelength range where NB filters cover is considerably limited.
This limitation inevitably makes LAE samples discrete with respect to redshift.
For spectroscopic samples, the limitation is somewhat relaxed, but it is still difficult to detect \lya{} emission at wavelengths obscured by the strong OH sky lines.

To understand the physical mechanism of \lya{} emission, previous studies have tried to find various correlations between \lya{} and the physical properties of host galaxies.
In general, \lya{} luminosity increases with higher star formation rate (SFR) and halo mass \citep{khostovan19}.
Typical cosmological simulations assign \lya{} luminosities using the relation between \lya{} and SFR or halo mass.
However, \lya{} emission varies due to various internal and surrounding effects.
\lya{} emission is sensitive to the existence of dust, and less dust galaxies tend to show \lya{} emission
\citep{Sobral18b, Santos20}.
Because UV slope $\beta$ is thought to reflect the amount of dust, a bluer UV slope can be indicative of \lya{} emission.
Younger age \citep{ArrabalHaro20} and lower mass \citep{khostovan19, Santos20} can be factors determining \lya{} emission.
LAEs are suggested to have smaller sizes compared to normal star-forming galaxies
\citep{Law12, Malhotra12, Paulino-Afonso18, Marchi18, Shibuya19, Ribeiro20}.
Smaller sizes can enhance the escape of ionizing photons. Following the positive correlation between ionizing photon escape fraction, $f_\mathrm{esc}$, and \lya{} photon escape fraction, $f_\mathrm{esc, Ly\alpha}$ \citep{Begley24, Maji22}, small galaxies tend to show \lya{} emission.
Some studies report that the spatial offset between the host galaxy and the \lya{} emission is a factor shaping \lya{} emission \citep{Lemaux21, Hoag19}.
\citet{Trainor16} argue that the velocity offset between the systemic redshift and \lya{} helps the escape of \lya{} photons.
\citet{Gazagnes20} suggest that H\,\textsc{i} covering fraction regulates the \lya{} emission.
Metallicity is also thought to be related to \lya{} emission \citep{Trainor16}.
Hard UV emission from low metallicity stellar population ionizes the neutral hydrogen, which enhances the creation and the escape of \lya{} photons.
Although much effort has been expended,
understanding which parameters determine \lya{} radiation is hampered by the complicated nature of \lya{} emission.

Previous studies try to predict \lya{} emission, using the correlation between \lya{} and the physical properties of host galaxies (\citealp{McCarron22, Ortiz23, Foran23, Napolitano23}, but see \citealp{Bolan24}).
However, the prediction using linear regression is not entirely accurate because \lya{} emission shows correlations to multiple galaxy properties and the variance of the correlation is large.
Some work uses multi-parametric analysis to predict \lya{} emission \citep{Runnholm20, Hayes23}.
Recent work begins to use machine learning techniques to deal with the multi-variable problem of the complex \lya{} emission.
\citet{McCarron22} use $k$-nearest method, and
\citet{Napolitano23} use random forest method to predict \lya{} emission from galaxies.

In this paper, we use a neural network to predict \lya{} emission from galaxy properties.
A neural network can handle non-linear relations between input parameters and an output value; therefore it is suitable for learning the complex relations between the physical properties and \lya{} emission.
By using the neural network as a prediction model, the presence or absence of \lya{} emission can be predicted from physical properties that can be estimated only by broad-band (BB) photometry.
We can efficiently detect LAEs in a wide field of view and in a wide and continuous redshift range, overcoming the limitations of spectroscopic observation or NB imagings.
Recently, JWST observation has started to provide a number of reionization-era galaxies.
The discovery of LAEs even at $z>7$, where H\,\textsc{i} IGM absorption should be intense, suggests the existence of ionized bubbles \citep{Jones23, Witstok24}.
The overwhelming increase in the number of LAEs predicted by the network will allow us to explore the evolution of the neutral fraction and the structure of reionization.

This paper is structured as follows.
Sec.~\ref{sec:data} describes the observational data and their analysis for the construction of a training dataset.
The training and evaluation of the prediction model are described in Sec.~\ref{sec:learning}.
In Sec.~\ref{sec:discussion}, we discuss the correlation between input parameters and the probability of \lya{} inferred by the model.
We apply the prediction model to galaxies detected by JWST observation and discuss the application to the study of reionization.

Throughout this paper, we use the AB magnitude system \citep{Oke83}.
We assume a $\Lambda$CDM cosmology with $h = 0.7$, $\Omega_\mathrm{m} = 0.3$, and $\Omega_\Lambda = 0.7$.
We use cMpc (pMpc) to indicate comoving (physical) scales.

\section{Data}\label{sec:data}

\begin{figure}
    \centering
    \includegraphics[]{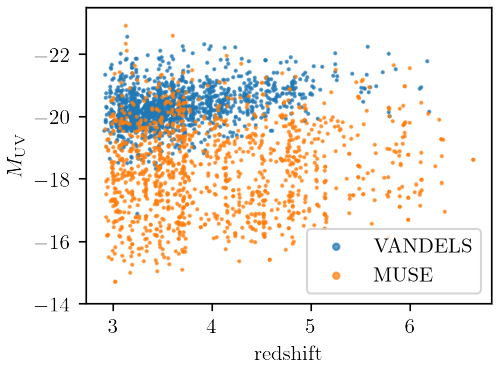}
    \caption{The relation between the redshift and $M_\mathrm{UV}$ of galaxies of VANDELS (blue) and MUSE (orange).
    $M_\mathrm{UV}$ is calculated using SED fitting (see Sec.\ref{sec:sed_fitting}).}\label{fig:z_muv}
\end{figure}

We collect spectroscopic data from the VANDELS \citep{McLure18} and MUSE \citep{Herenz17, Bacon17} spectroscopic surveys.
The VANDELS survey targets the UDS (Ultra Deep Survey) and CDFS (Chandra Deep Field South) fields.
The target selection is based on the photometric redshift of star-forming galaxies with $i < 27$.
The survey is carried out using VIMOS on VLT.
The wavelength range of the spectra is $4800\,\mathrm{\AA} < \lambda < 9800\,\mathrm{\AA}$, which captures the \lya{} emission at $2.9 < z < 7.0$.
The final data release of VANDELS \citep{Garilli21} contains 2087 galaxies at $1 < z < 6.5$.

The MUSE survey is carried out in the GOODS-S (CDFS) and COSMOS fields using MUSE on VLT.
MUSE survey is comprised of two different depth observations.
MUSE-Wide \citep{Herenz17, Urrutia19} covers the COSMOS and GOODS-S fields,
and MUSE-Deep \citep{Bacon17, Inami17, Bacon23} targets HUDF (Hubble Ultra-Deep Field) located at the center of the GOODS-S field.
A 1$\sigma$ emission line
detection sensitivity of MUSE-Wide data is $1 \times 10^{-19}\,\mathrm{erg\,s^{-1}\,cm^{-2}\,\AA^{-1}}$,
and that of MUSE-Deep data is $5.5 \times 10^{-20}\,\mathrm{erg\,s^{-1}\,cm^{-2}\,\AA^{-1}}$.
The wavelength range of the spectra is $4800\,\mathrm{\AA} < \lambda < 9300\,\mathrm{\AA}$.
Object detection relies on the presence of emission lines,
leading to select low-mass galaxies.
Therefore, the MUSE survey complements the galaxy of the VANDELS survey, which is basically a magnitude-limited sample.
\citet{Kerutt22} determine the \lya{} measurement of LAEs detected with the MUSE-Wide and MUSE-Deep surveys.
\citet{Schmidt21} provide a catalogue of 2052 galaxies at $1.5 < z < 6.4$.
Fig.~\ref{fig:z_muv} shows the relation between the redshift and $M_\mathrm{UV}$ of the VANDELS and MUSE samples.

\subsection{\lya{} Flux Measurement}\label{sec:flux_measurement}
We measure the \lya{} flux of spectroscopically observed galaxies from VANDELS and MUSE at $3 \lesssim z \lesssim 6$.
In this redshift range, we assume the reionization is completed and the attenuation of \lya{} emission by the global neutral fraction is negligible.
For VANDELS samples, we use the spectroscopic data to measure the \lya{} flux and equivalent width (EW), which are not included in the catalogue of \citet{Garilli21}.
First, we fit the continuum redward of \lya{} in the wavelength range of $1250\,\mathrm{\AA} < \lambda < 2500 \,\mathrm{\AA}$ in the rest frame with a power law with an index of $\alpha$ ($A\lambda^\alpha$).
The UV continuum is then subtracted using a fitted power function.
We determine the \lya{} line flux by directly integrating the continuum-subtracted spectra over $1215.67 \pm 3\,\mathrm{\AA}$.
The \lya{} EW is derived by dividing the \lya{} flux by the UV continuum flux at the position of \lya{} ($\lambda = 1215.67\,\mathrm{\AA}$), which is calculated using the result of power-law fitting to the redward of \lya{} as
\begin{align}
    f_\mathrm{UV} = A \times (1215.67\times(1+z))^\alpha,
\end{align}
where $A$ and $\alpha$ are the fitting parameters.
For strong emitters ($\mathrm{EW}_0 > 10\,\mathrm{\AA}$), which often have \lya{} emission extending beyond $\pm 3 \,\mathrm{\AA}$, we determine the integration window by fitting the \lya{} emission line with a Gaussian function.
When the \lya{} $\mathrm{EW}_0$ calculated over $1215.67 \pm 3\,\mathrm{\AA}$ is larger than $10\,\mathrm{\AA}$, we recalculate the \lya{} flux by directly integrating the observed flux over $\pm 3\,\sigma$ around the central wavelength of the \lya{} emission.
Because the \lya{} emission often has an asymmetric profile or double-peaked profile,
we use the directly integrated flux instead of the total area of the fitted Gaussian function.
The typical Gaussian $\sigma$ is $\sim 1\,\mathrm{\AA}$.
We assume the \lya{} flux extending out of the $\pm 3\sigma$ range is negligible.

To calculate the error of \lya{} flux, we use Monte Carlo simulation.
We randomly add Gaussian error, extracted from the noise spectrum, to each wavelength bin of the spectrum.
We generate 1000 noise-added spectra
and calculate \lya{} flux of all the spectra in the same way as described above.
The flux and 1$\sigma$ error are determined by the median and 16--84th percentile of the distribution of flux measurement for the 1000 spectra, respectively.
We visually inspect the spectra and remove objects whose \lya{} are contaminated by severe noise or sky emission lines.

\citet{Talia23} calculate EW of VANDELS galaxies by fitting with a model spectrum that considers stellar continuum and emission and absorption lines.
EW measurements are mostly consistent with this study for the objects with positive EW values (i.e., \lya{} detected objects).
There is some systematic offset for the negative EW values because we do not deal with \lya{} absorption.
This does not impact our study because we are only interested in whether a galaxy emits \lya{} line or not.
As described in Sec.~\ref{sec:learning}, we select LAEs based on visual inspection.

We use the \lya{} flux and EW of the MUSE sample measured in \citet{Kerutt22}.
The measurement procedures of \citet{Kerutt22} are as follows.
\lya{} emission lines are fitted using an asymmetric Gaussian function.
To calculate EW, continuum flux is estimated using HST bands.
UV continuum slope is taken into account by using two HST bands if possible.
The UV continuum is assumed to be a power law function ($f (\lambda) \propto \lambda^\alpha$).

The flux calibration of VANDELS spectra is carried out using $i$-band photometry.
This corrects for the flux loss in VANDELS spectra.
The measurement of \lya{} flux using MUSE IFU can be performed beyond the aperture used in photometry.
It leads to extracting an extended component of \lya{} emission whereas VANDELS observation takes into account the \lya{} flux from the core component.
To calibrate this difference, we compare 25 objects appearing in both the VANDELS and MUSE catalogues.
The median and 16th-84th percentile of the ratio of \lya{} flux between VANDELS and MUSE is $0.55^{+0.56}_{-0.31}$.
We regard this value as the correction factor and multiply 0.55 by the \lya{} flux and EW of MUSE galaxies.

\subsection{Broad-band photometry}\label{sec:matching}
In addition to the spectroscopic measurements, we collect BB photometry to obtain the physical properties of the sample.
BB photometry of VANDELS galaxies is included in VANDELS catalogue \citep{Garilli21}.
Near-UV to near-IR photometry is observed with HST and Spitzer/IRAC in the central regions of UDS and CDFS.
Photometric data for the regions that are not covered by HST and Spitzer/IRAC are taken from ground-based observations, such as CFHT, Subaru, VISTA, and UKIRT.

Since the BB photometry of the MUSE sample is not included in the MUSE catalogue, they are extracted from CANDELS \citep{Guo13}, 3D-HST \citep{Skelton14}, and UVUDF \citep{Rafelski15}.
Cross matching of the sources and UV continuum counterparts is performed in \citet{Kerutt22}.
Following their results, we obtain 720 objects matched in CANDELS, 354 in 3D-HST, and 364 in UVUDF after excluding duplication among these surveys.
Tab.~\ref{tab:dataset} summarizes the number of galaxies and the list of available filters for the VANDELS and MUSE surveys.

\begin{table*}
    \centering
    \caption{The overview of the number of galaxies and the list of available filters for the VANDELS and MUSE surveys.}\label{tab:dataset}
    \begin{tabular}{lcccccccc}
        \hline
        Spectroscopic survey & \multicolumn{4}{c}{VANDELS} & \multicolumn{4}{c}{MUSE} \\
        Target field & \multicolumn{2}{c}{CDFS} & \multicolumn{2}{c}{UDS} & GOODS-S & GOODS-S & COSMOS & GOODS-S \\
        Photometry & HST & Ground & HST & Ground & CANDELS & 3D-HST & 3D-HST & UVUDF \\
        Number of galaxies & 410 & 335 & 397 & 355 & 720 & 65 & 289 & 364 \\
        \hline
        Available filters & $U$ & $U$ & $u$ & $u$ & $U$ & $U$ & $u$ & F435W\\
        & F435W & $B$ & $B$ & $B$ & F435W & F435W & $B$ & F606W\\
        & F606W & IA484 & $V$ & $V$ & F606W & $B$ & $g$ & F755W\\
        & F755W & IA527 & F606W & $r$ & F755W & $V$ & $V$ & F850LP\\
        & F814W & F606W & $r$ & $i$ & F814W & F606W & F606W & F105W\\
        & F850LP & IA624 & $i$ & $z$ & F850LP & $R$ & $r$ & F125W\\
        & F098M & $R_c$ & F814W & $z^{++}$ & F098M & $R_c$ & $i$ & F140W\\
        & F105W & IA679 & $z$ & $Y$ & F105W & F755W & F814W & F160W\\
        & F125W & IA738 & $Y$ & $J$ & F125W & $I$ & $z$ & \\
        & F160W & IA767 & F125W & $H$ & F160W & F850LP & $z^{++}$ & \\
        & $K_s$ & $Z$ & $J$ & $K$ & $K_s$ & F125W & $Y$ & \\
        & IRAC1 & F850LP & F160W & IRAC1 & IRAC1 & $J$ & F125W & \\
        & IRAC2 & $Y$ & $H$ & IRAC2 & IRAC2 & F140W & $J$ & \\
        &  & $J$ & $K_s$ &  & IRAC3 & F160W & F140W & \\
        &  & $H$ & $K$ &  & IRAC4 & $H$ & F160W & \\
        &  & $K_s$ & IRAC1 &  &  & $K_s$ & $H$ & \\
        & & IRAC1 & IRAC2 &  &  & IRAC1 & $K_s$ & \\
        &  & IRAC2 &  &  &  & IRAC2 & IRAC1 & \\
        &  &  &  &  &  & IRAC3 & IRAC2 & \\
        &  &  &  &  &  & IRAC4 & IRAC3 & \\
        &  &  &  &  &  &  & IRAC4 & \\
        \hline
    \end{tabular}
\end{table*}

\subsection{SED Fitting}\label{sec:sed_fitting}
\begin{figure*}
    \centering
    \includegraphics[]{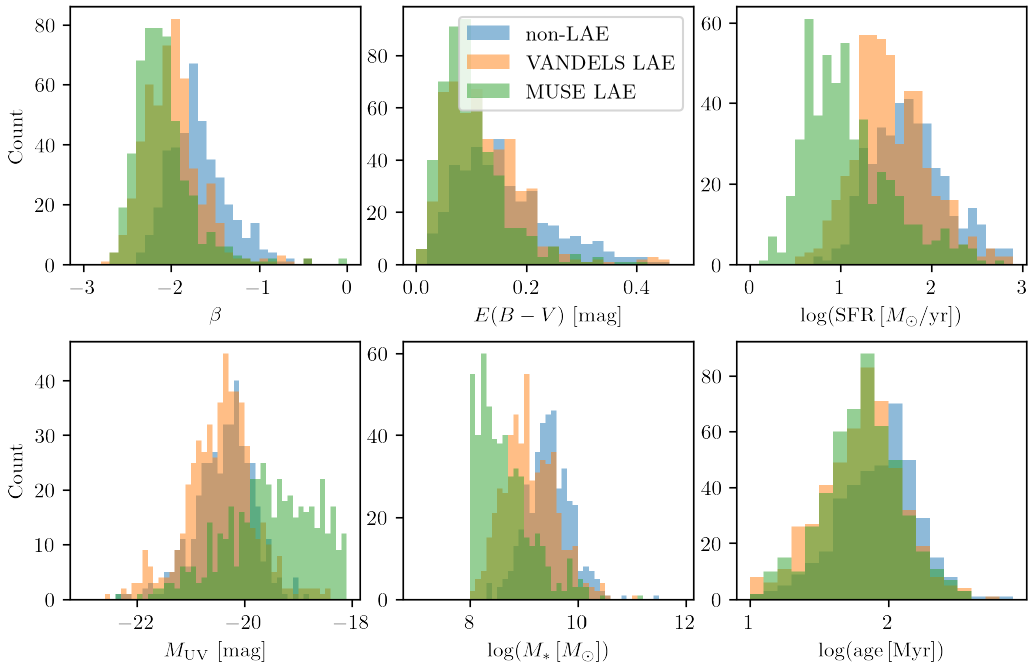}
    \caption{The distribution of six physical properties ($\beta$, $E(B-V)$, SFR, $M_\mathrm{UV}$, $M_*$, and age). Blue, orange, and green histograms show non-LAEs from VANDELS, LAEs from VANDELS, and LAEs from MUSE, respectively. Galaxies taken from MUSE are all classified as LAEs because MUSE catalog selects objects based on line detections.}\label{fig:param_hist}
\end{figure*}

\begin{figure*}
    \centering
    \includegraphics[]{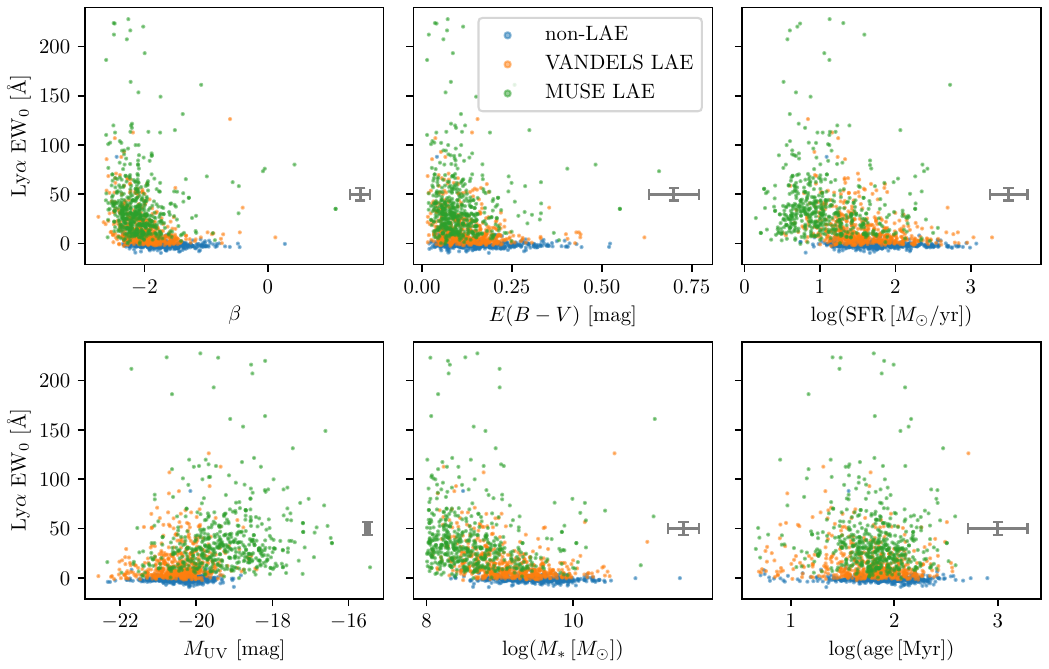}
    \caption{The relation between the EW$_0$ of \lya{} and six physical properties ($\beta$, $E(B-V)$, SFR, $M_\mathrm{UV}$, $M_*$, and age). Colors are the same as Fig.~\ref{fig:param_hist}. The error bar at the bottom right on each panel shows the typical error of each parameter and EW$_0$.}\label{fig:param_ew}
\end{figure*}

We estimate the galaxy properties using SED fitting code \texttt{CIGALE} \citep{Boquien19}.
It assumes the single stellar population model of \citet{Bruzual03} and the Chabrier initial mass function \citep{Chabrier03}.
We adopt $\tau$-model star formation history with $\tau = 10\mbox{--}120\,\mathrm{Myr}$.
The searching range of model parameters are age with $4\mbox{--}1600\,\mathrm{Myr}$, metallicity with $0.05\mbox{--}1Z_\odot$, dust attenuation with $0.0 < E(B-V) < 1.0$, and the ionization parameter with $-4 < \log U < -2$.
The dust attenuation curve assumes Calzetti extinction law \citep{Calzetti00}.
Redshift is fixed at the spectroscopic redshift.
The spectroscopic redshift is taken from the catalogue of \citet{Garilli21} for the VANDELS sample and \citet{Schmidt21} for the MUSE sample.
For more information about the spectroscopic redshift measurement, see \citet{Pentericci18} and \citet{Schmidt21} for VANDELS and MUSE, respectively.

We remove galaxies that fail SED fitting from the sample.
To make sure SED fitting is correctly performed, we exclude galaxies with reduced $\chi^2 > 2.5$.
While the redshift of the sample is spectroscopically confirmed, we impose that the measurement of photometric redshift is correctly determined.
When we apply the prediction model to galaxies that have no spectroscopic observation, we use photometric redshift for SED fitting.
In Sec.~\ref{sec:cosmos_sc4k_data}, we fix the redshift at the photometric redshift during SED fitting.
If photometric redshift measurement is poorly determined, SED fitting gives a wrong answer.
It is impossible to accurately infer LAEs using the prediction model with the wrong SED fitting.
For this reason, we require $\Delta z = |z_\mathrm{phot} - z_\mathrm{spec}|/(1+z) < 0.15$ to ensure the accuracy of photometric redshift from the catalogs of VANDELS, CANDELS, 3D-HST, and UVUDF.
As shown in Fig.~\ref{fig:param_hist}, most of the faint, lower-mass objects come from the MUSE survey, and they are all classified as LAEs.
This is because the MUSE survey is based on emission line detection while the VANDELS survey is a mass-limited sample.
To mitigate the impact of this difference on the prediction model, we remove objects with $M_* < 10^8 M_\odot$.
AGNs that are identified by broad emission lines in the VANDELS Survey \citep{Garilli21} are excluded from the final sample.
For the MUSE sample, we remove two Type 2 QSOs that are mentioned in \citet{Urrutia19}.
The number of galaxies for the training dataset is 926 from VANDELS and 507 from MUSE, respectively.

\subsection{COSMOS2020 and SC4K samples}\label{sec:cosmos_sc4k_data}
To validate the trained model with galaxies that have no spectroscopic measurement, we use two photometric samples: COSMOS2020 and SC4K.
COSMOS2020 \citep{Weaver21} provides a deep multi-band catalogue of galaxies.
COSMOS2020 derived photometric redshift using UV to IR photometry.
It is expected that the majority of the galaxies in the COSMOS2020 catalogue are non-LAEs.

SC4K \citep{Sobral18} is an NB and medium-band (MB) selected LAE catalogue.
SC4K uses 16 filters to identify $\sim4000$ LAEs in the COSMOS field at several discrete redshift ranges at $2 < z < 6$.
To estimate the physical properties of SC4K LAEs, we obtain BB photometry by cross-matching with the COSMOS2020 catalogue.
The number of SC4K LAEs matched in the COSMOS2020 catalogue is 3453.

For a fair comparison of the two catalogues, we select galaxies whose photometric redshift from the COSMOS2020 catalogue is included in the redshift ranges of SC4K LAEs.
When cross-matching the SC4K sample and the COSMOS2020 sample, LAE candidates of SC4K are only 1.5\% of galaxies of COSMOS2020 within the redshift ranges of the SC4K LAEs.
Note that LAE candidates of SC4K, which indicate a flux excess in NB or MB, have large EWs with $\mathrm{EW}_0>25\,\mathrm{\AA}$ \citep{Sobral18}, while
The COSMOS2020 sample may include LAEs with smaller EWs.

The physical properties of COSMOS and SC4K galaxies are estimated by SED fitting using \texttt{CIGALE} in the same way as in Sec.~\ref{sec:sed_fitting}.
Because spectroscopic redshift is not available, we fix the redshift at the photometric redshift for COSMOS2020 galaxies.
The redshift of SC4K LAEs is fixed at the central wavelength of NB/MB corresponding to a \lya{} emission.

As COSMOS2020 has a wide variety of galaxies, we select galaxies based on the same selection criteria as in the training dataset.
The criteria are the number of photometries with $S/N > 5$ is larger than or equal to ten, $-22 < M_\mathrm{UV} < -18$, $M_* > 10^8M_\odot$, and reduced $\chi^2 < 2.5$.
We apply the criteria of $-22 < M_\mathrm{UV} < -18$, $M_* > 10^8M_\odot$, and reduced $\chi^2 < 2.5$ to the SC4K sample because they are detected in the fewer number of filters compared to COSMOS2020.
The numbers of galaxies are 67068 and 2273 from COSMOS2020 and SC4K, respectively.

\subsection{JWST Sample}\label{sec:jwst_data}
We apply the prediction model to galaxies detected by JWST observation in Sec.~\ref{sec:JWST}.
We use public JWST imaging data in of CEERS \citep[ERS-1345;][]{Finkelstein23}, the COSMOS and UDS fields from PRIMER \citep[GO-1837;][]{Dunlop21}, and the GOODS-N/-S fields from FRESCO \citep[][]{Oesch23} and GO-1963 \citep[][]{Williams21}.
The photometry is performed with \texttt{Grizli} \citep{Brammer23} for HST and JWST imagings.
The images and multiband catalogues are available online\footnote{\url{https://dawn-cph.github.io/dja/index.html}}.
We use the multiband catalogue of the public release version 6\footnote{\url{https://s3.amazonaws.com/grizli-v2/JwstMosaics/v6/index.html}}.

SED fitting is carried out in the same way as in Sec.~\ref{sec:sed_fitting} using \texttt{CIGALE} to estimate the physical properties of the galaxies.
The redshift is fixed at the best-fitting photometric redshift determined by \texttt{EAZY} \citep{Brammer08}.

Similar to Sec.~\ref{sec:cosmos_sc4k_data}, we select objects with $-22 < M_\mathrm{UV} < -18$, $10^8 < M_* / M_\odot < 10^{11}$, reduced $\chi^2 < 2.5$, and redshift range of $2 < z < 6$.
Galaxies with calculated stellar mass of $M_* > 10^{11}$ are excluded because the SED fitting is poorly determined.
Additionally, we exclude galaxies that have flux 3$\sigma$ higher than the model spectrum calculated by CIGALE at the wavelength shorter than \lya{} to assure the accuracy of photometric redshift.
The number of galaxies satisfying the condition above is 6938.
The median value of the stellar mass of the JWST sample is $10^{8.5}\,M_\odot$.
The JWST sample contains less massive galaxies than the training dataset with a median stellar mass of $10^{9.1}\,M_\odot$, and COSMOS2020 with $10^{9.3}\,M_\odot$.

\section{Prediction of LAE from physical properties}\label{sec:learning}
\subsection{Training Dataset}\label{sec:training_data}
We construct a prediction model using the training dataset consisting of galaxies from VANDELS and MUSE.
For the prediction of LAEs, we use six parameters derived from SED fitting (SFR, stellar mass $M_*$, UV absolute magnitude $M_\mathrm{UV}$, age, UV slope $\beta$, and dust attenuation $E(B-V)$).
We use these six parameters because they are thought to be related to \lya{} emission and can be estimated by SED fitting with BB photometries.
To train the model, we categorize the VANDELS data into two labels: LAE and non-LAE.
LAEs are selected based on visual inspection.
It may seem better to set a certain threshold of EW to classify the two like LAE surveys using the NB technique.
However, LAEs detected with NB techniques tend to be biased towards LAEs with high \lya{} EW.
Since our galaxy sample is based on spectroscopic data, we can use the detailed analysis of \lya{} flux as described in Sec.~\ref{sec:data}.
The measurement reveals a number of galaxies with faint ($\mathrm{EW}_0 < 10\,\mathrm{\AA}$) but clear \lya{} emission lines.
On the other hand, despite being a non-LAE, it sometimes shows $\mathrm{EW}_0 \sim 5\,\mathrm{\AA}$ due to uncertainty of the spectrum.
Thus, selecting LAEs with an EW threshold may miss these faint LAEs or suffer from contaminants.
Furthermore, some of the galaxies have both emission and absorption at the position of \lya{} line.
It makes difficult the precise measurements of \lya{} EW because uncertainties in the EW measurement increase when EW decreases.
For these reasons, we rely on visual inspection to select LAEs.
As a result, the galaxies with $\mathrm{EW}_0 > 7\,\mathrm{\AA}$ are all classified as LAEs by our visual inspection.
The boundary between LAE and non-LAE roughly corresponds to $\mathrm{EW}_0=3\,\mathrm{\AA}$.
We confirmed that non-LAE is approximately distributed around $\mathrm{EW}_0=0\,\mathrm{\AA}$.
Since the MUSE sample is originally selected by the detection of \lya{} emission lines, all galaxies are classified as LAEs.
The final dataset consists of 520 LAEs and 406 non-LAEs from VANDELS and 507 LAEs from MUSE.
The dataset is split into an 80\% training sample and a 20\% test sample.

Fig.~\ref{fig:param_hist} shows the distributions of the six physical properties,
and Fig.~\ref{fig:param_ew} shows the relation between EW and the six physical properties.
As shown by previous studies \citep{Sobral18b, Santos20, ArrabalHaro20, khostovan19}, LAEs tend to have lower SFR, lower mass, fainter UV magnitude, younger age, bluer UV slope, and less dust content.

\subsection{Neural Network}\label{sec:nn}
\begin{figure*}
    \centering
    \includegraphics[]{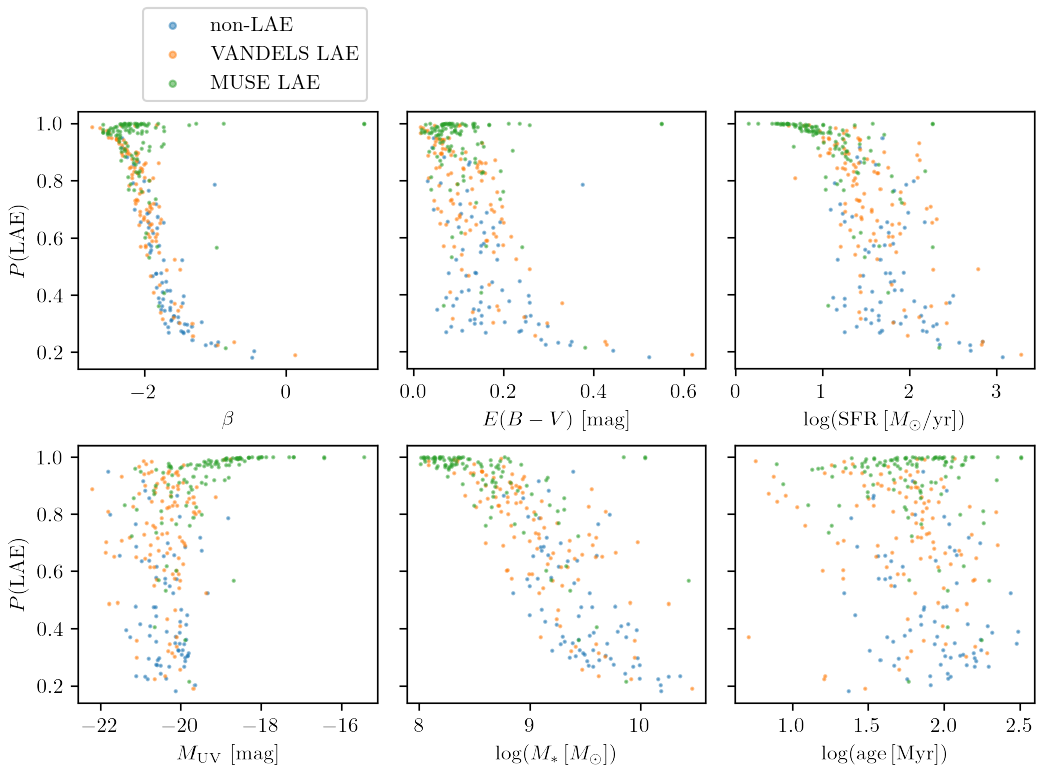}
    \caption{The relation between the output probability of $P(\mathrm{LAE})$ and the six input parameters for VANDELS LAEs (orange), MUSE LAEs (green), and non-LAEs (blue). }\label{fig:param_plae}
\end{figure*}

\begin{figure}
    \centering
    \includegraphics[]{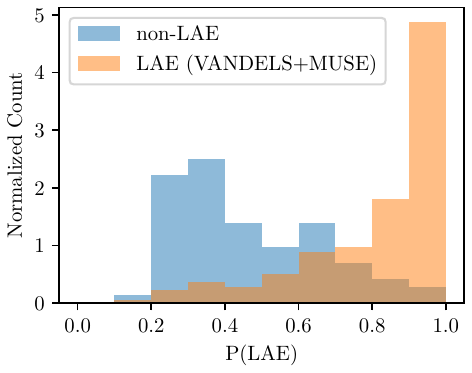}
    \caption{The distribution of output probability of $P(\mathrm{LAE})$ for LAEs (orange), and non-LAEs (blue). }\label{fig:plae}
\end{figure}

We use a neural network to predict the probability that a galaxy exhibits \lya{} emission lines.
The input parameters are six physical properties derived from SED fitting (SFR, $M_*$, $M_\mathrm{UV}$, age, $\beta$, and $E(B-V)$).

We construct a neural network architecture using TensorFlow/Keras \citep{chollet2015keras, tensorflow}.
The neural network consists of five hidden layers with 64 nodes per layer.
To determine the optimal number of hidden layers, we test several neural network architectures changing the number of hidden layers.
We find that the number of hidden layers smaller than four does not have enough capacity to separate LAEs and non-LAEs.
On the other hand, the number of hidden layers larger than five does not improve the performance.
We use eLU with a parameter $\alpha=0$ as an activation function after each hidden layer.
We apply 25\% dropout connection after each hidden layer to prevent overfitting.
The final layer of the neural network is a dense layer with a single node, using a sigmoid activation function to ensure the output values are between 0 and 1.
The learning rate sets 0.001 as an initial value,
decaying by a factor of 0.96 every 100 steps.
We use Adam optimizer \citep{Kingma15} and binary cross entropy loss.
Training proceeds with a maximum of 400 epochs, but it ends at the point that gives the minimum validation loss to avoid overfitting.

During the training, we employ 5-fold cross-validation
in which the training sample is divided into five parts, each with 20\%.
We train the network five times, each time using a different 20\% part as the validation set and the remaining 80\% part for training.
After that, we take the average of the outputs of five networks.
This ensures that our network does not overfit to any specific subset of the data.

Besides, we also employ a Monte Carlo approach to account for the uncertainty in the input physical properties.
We randomly add Gaussian errors to the six input parameters.
The standard deviation of the Gaussian error is set to the uncertainties of the parameters from the CIGALE output.
For each run, the network is trained with an uncertainty-added dataset in the same way as described above.
We repeat this procedure ten times.
The sample splitting of the 5-fold cross-validation is fixed in the whole process.
The output score is an average of the outputs of ten uncertainty-added models.
We regard this score as the probability of LAE, $P(\mathrm{LAE})$, though it is not strictly a probability.
This ensemble approach helps improve the robustness of our predictions and ensures our model takes into account the uncertainty in the measured physical properties.

\subsection{Validation}\label{sec:validation}
\begin{figure}
    \centering
    \includegraphics[]{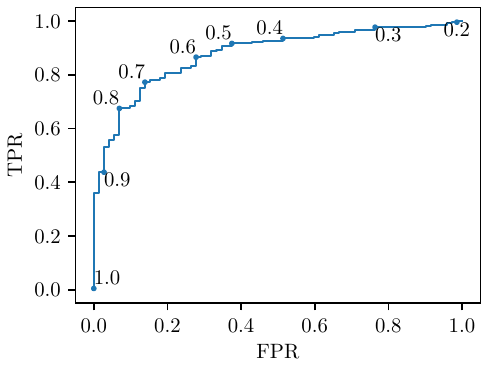}
    \caption{The ROC curve of the trained model. True Positive Rate (TPR) and False Positive Rate (FPR) are defined as $\mathrm{TPR} = \mathrm{TP} / (\mathrm{TP} + \mathrm{FN})$ and $\mathrm{FPR} = \mathrm{FP} / (\mathrm{TN} + \mathrm{FP})$, where TP, FN, TP, and TN are true positive, false negative, true positive, and true negative, respectively. The value along the curve indicates the threshold of $P(\mathrm{LAE})$.}\label{fig:roc_pr}
\end{figure}

Here, we check how well the network performs with the remaining 20\% of the test sample that we do not use for training.
Fig.~\ref{fig:param_plae} shows the relation between the output $P(\mathrm{LAE})$ of the trained network and the input physical parameters.
Similar to Fig.~\ref{fig:param_ew}, $P({\mathrm{LAE}})$ has strong correlation with $\beta$ and $M_*$ and moderate correlation with SFR, $M_\mathrm{UV}$.
Age and $E(B-V)$ are less sensitive to $P(\mathrm{LAE})$.
Fig.~\ref{fig:plae} shows the distribution of the output probability of LAEs of the test sample.
The LAEs in the test sample tend to have a higher probability, while non-LAEs show the opposite trend.
As shown in Fig.~\ref{fig:plae}, no galaxy have $P(\mathrm{LAE}) < 0.1$.
This is because the parameter distributions of non-LAEs typically overlap with those of LAEs.
There exist LAEs with $\beta > -2$ or $E(B-V) > 0.2$ where non-LAEs are dominant.
As a result, none of the input parameters is a distinctive indicator of non-LAEs, and no galaxy has $P(\mathrm{LAE}) < 0.1$.
On the other hand, a large fraction of LAEs have $P(\mathrm{LAE}) > 0.9$.
This is because the training dataset does not include low-mass non-LAEs, so almost all of the galaxies with $M_\mathrm{UV} > -19$ or $M_* < 10^9\, M_\odot$ are classified as LAEs.
In such case, the network can preferentially predict the low mass galaxies as LAE with $P(\mathrm{LAE}) > 0.9$.
Fig.~\ref{fig:roc_pr} shows the ROC curve of the trained model.
When we define LAEs as $P(\mathrm{LAE}) > 0.7$, the predicted LAE sample has 77\% true positive rate (TPR) and 14\% false positive rate (FPR).
Similarly, for the threshold of $P(\mathrm{LAE}) > 0.5$, TPR is 91\%, and FPR is 38\%,
and for the threshold of $P(\mathrm{LAE}) > 0.8$, TPR is 67\%, and FPR is 7\%.
The threshold of $P(\mathrm{LAE}) > 0.8$ is preferable when the purity of a sample is important, while $P(\mathrm{LAE}) > 0.5$ is suitable for reducing missed detections despite potential contamination.
It should be noted that there is no $z$-dependence of these values and the ROC curve does not change significantly with redshift.
The area under the ROC curve (AUC) is 0.88.
If we select LAEs based only on UV slope $\beta$, the AUC is 0.82.
Similarly, when selecting LAEs based on one of the parameters, $M_*$ and SFR, the AUCs are 0.84 and 0.77, respectively.
This demonstrates the advantage of the neural network, which handles multiple variables compared to the selection based on a single variable.

As described above, the MUSE sample occupies the fainter parameter space, all of which are classified as LAEs.
To test the impact of adding the MUSE sample, we train the network with only the VANDELS sample.
The AUC does not change when we classify the VANDELS LAEs and non-LAEs using the model.
Therefore, the addition of the MUSE sample does not degrade the performance of the model.
Adding the MUSE sample extends the range of the input parameters, in particular fainter objects, as shown in Fig.~\ref{fig:param_hist} without sacrificing the performance of the prediction model.
$P(\mathrm{LAE})$ is found not to correlate with \lya{} flux or EW. 
Therefore, unfortunately, it seems difficult to predict \lya{} flux and EW with this model.

\citet{Napolitano23} detect LAEs using a random forest classifier.
Fig. 10 of \citet{Napolitano23} reports the total number of TP, FP, TN, and FN for their test sample.
Following the result, the TPR is 65\%, and the FPR is 9\%.
This result is similar to our result when adopting the threshold of $P(\mathrm{LAE}) > 0.8$.
This shows that a neural network and a random forest exhibit similar performance while various techniques are available for this kind of analysis.
In this study, we select a neural network due to its robust performance and ability to handle complex data structures.

\section{Discussion}\label{sec:discussion}

\subsection{Feature Importance}\label{sec:feature_importance}

\begin{figure}
    \centering
    \includegraphics[]{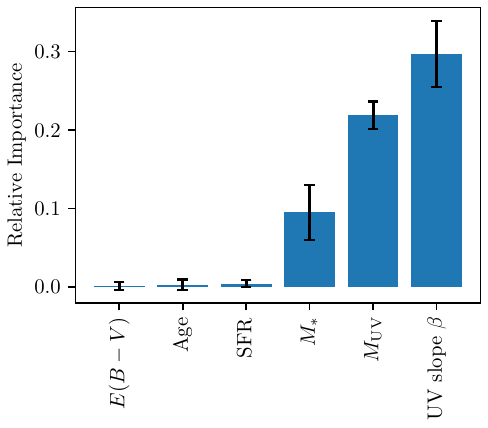}
    \caption{Relative importance of the six parameters estimated by permutation feature importance. The error bar shows the standard deviation of values derived for 50 networks (5-fold cross validation $\times$ 10 Monte Carlo simulations).}\label{fig:feature_importance}
\end{figure}

We discuss which input parameters have the most strong influence on the determination of $P(\mathrm{LAE})$.
The permutation feature importance \citep[PFI, ][]{Altmann10} is one of the methods to evaluate the relative importance of the input parameters, which is derived by shuffling the values of a specific input variable in the dataset and assessing its impact on the model's performance.
When an input parameter is important to the model output, model performance gets worse significantly after shuffling the values.
On the other hand, a parameter is less important if the model performance does not vary much.
Therefore, the difference can be regarded as the importance of the input parameter.
Fig.~\ref{fig:feature_importance} shows the importance of the input parameters derived by the permutation feature importance method.
$\beta$, $M_\mathrm{UV}$, and $M_*$ are found to significantly affect the model's output, as inferred by the correlation shown in Fig.~\ref{fig:param_plae}.
On the other hand, SFR, $E(B-V)$, and age show the importance consistent with zero.
However, the permutation feature importance method can provide misleading results for correlated features because the model can infer from other correlated parameters even when shuffling values of one parameter.
Therefore, low importance does not necessarily mean that the parameter is not relevant to the model output when the parameter is correlated with other parameters.
The other possibility is that the parameters with the importance consistent with zero have another correlated parameter more strongly connected to the model output.
As inferred by the physical connection with $\beta$, this correlation between $\beta$ and $E(B-V)$ can degrade the measurement of importance of $E(B-V)$.
Similarly, SFR has a strong correlation with $M_*$, and the importance of SFR might be underestimated.
The age does not show an apparent correlation with the other parameters; therefore, we interpret that the age does not impact the model output.
It might also be due to the age estimate of SED fitting that has a large uncertainty as shown in Fig.~\ref{fig:param_ew}.

\subsection{Misclassified objects}\label{sec:misclassify}
UV slope $\beta$ is the primary factor to determine $P(\mathrm{LAE})$; bluer UV slope leads to the emergence of \lya{} emission.
However, the training dataset contains a distinct population of LAEs with a red UV slope.
The prediction model tends to miss these red LAEs, leading to false negatives.
Several observational studies suggest the existence of old LAEs, with ages significantly above 100 Myr \citep{Finkelstein09, Pentericci09, Iani24}.
\citet{Shapley01} proposed two distinctive old and young populations in \lya{} bright phases: a galaxy appears as an LAE (blue LAE) during its initial starburst epoch when it is still dust-free, then it becomes a dusty Lyman break galaxy (LBG) having Ly$\alpha$ absorption after ISM enrichment.
\lya{} emission could appear again as red LAE when a galaxy becomes less dusty, at least on the line of sights, due to an outflow when it is older than a few $\times10^8\,\mathrm{yr}$.
Furthermore, understanding the physical mechanism of red LAEs and adopting other parameters that are also sensitive to \lya{} emission are required to correctly detect red LAEs.
As shown in Fig.~\ref{fig:param_plae}, there is a strong correlation between the probability of LAEs and the stellar mass, with low-mass galaxies showing higher probabilities of LAEs.
This result seems to be reasonable given that lower mass galaxies tend to emit \lya{}.
However, this correlation contributes to false positives because all low-mass galaxies do not necessarily emit \lya{}.
Besides, the prediction model might be biased toward misclassified low-mass galaxies because all lower-mass galaxies of the training dataset are classified as LAEs as shown in Fig.~\ref{fig:param_plae}.
Even if a galaxy has a blue UV continuum, \lya{} emission can be obscured due to a viewing angle effect.
In a patchy ISM scenario, an ionized channel that opens towards the line of sight boosts the escape of \lya{} photons, but otherwise, \lya{} photons are scattered by a neutral cloud \citep{Smith19}.

Galaxies with the most pronounced \lya{} emission generally have the smallest \lya{} velocity offsets \citep[e.g.,][]{Tang24}.
This may reflect lower H\,\textsc{i} column densities that allow \lya{} to escape without diffusing to large velocities.
Furthermore, introducing a factor related to H\,\textsc{i} content, such as size \citep{Pucha22}, may improve the model performance.
As described in Sec.~\ref{sec:data}, we added low-mass LAEs from the MUSE survey to complement the VANDELS sample.
This sample bias can influence the prediction results as discussed above.
However, our prediction model realizes the classification of LAEs for at least moderate-mass galaxies with the current limited dataset.
For the more accurate prediction of low-mass LAEs, we need to construct a prediction model trained with an unbiased dataset regarding stellar mass using deeper observations.
While adding the MUSE sample to the training dataset does not degrade the classification of moderate-mass LAEs mainly from the VANDELS survey as shown in Sec.~\ref{sec:validation}, including even lower mass non-LAEs from e.g., JWST surveys will enhance the model's performance, especially for classification of low-mass galaxies.

\subsection{Validation with other galaxy survey data}\label{sec:cosmos_sc4k}

\begin{figure}[!h]
    \centering
    \includegraphics[]{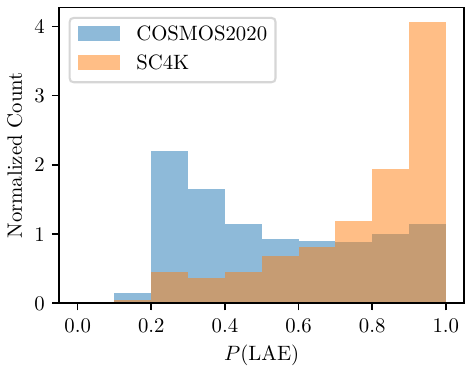}
    \caption{The distribution of $P(\mathrm{LAE})$ for COSMOS2020 (blue) and SC4K (orange) inferred by the prediction model.}\label{fig:cosmos_sc4k}
\end{figure}

To test whether the trained model is applied to other observational samples (especially samples without spectroscopy), we apply the model to two publicly available galaxy samples, COSMOS2020 and SC4K.
We calculate $P(\mathrm{LAE})$ for COSMOS2020 and SC4K galaxies using the trained neural network.
Fig.~\ref{fig:cosmos_sc4k} shows the distribution of $P(\mathrm{LAE})$ of COSMOS2020 and SC4K galaxies.
As expected, SC4K LAEs indicate higher $P(\mathrm{LAE})$ while COSMOS2020 galaxies have a lower probability.
This result shows that the neural network is applicable to galaxies where spectroscopic redshift is not available.
Assuming all of the SC4K galaxies are LAEs, the TPR is 72\% when selecting LAEs with $P(\mathrm{LAE}) > 0.7$.
Note that SC4K selects LAE candidates, which may contain contaminants.
While $P(\mathrm{LAE})$ distribution of the COSMOS2020 galaxies is biased towards non-LAEs as expected, it has relatively more galaxies that show $P(\mathrm{LAE})>0.7$ than in Fig.~\ref{fig:plae} even though many of them are expected to be non-LAE.
This may attributed to the fact that COSMOS2020 contains a small portion of LAEs, especially those with $\mathrm{EW}_0 < 25\,\mathrm{\AA}$.

\subsection{Application to JWST Photometric Data}\label{sec:JWST}

\begin{figure*}
    \centering
    \includegraphics[]{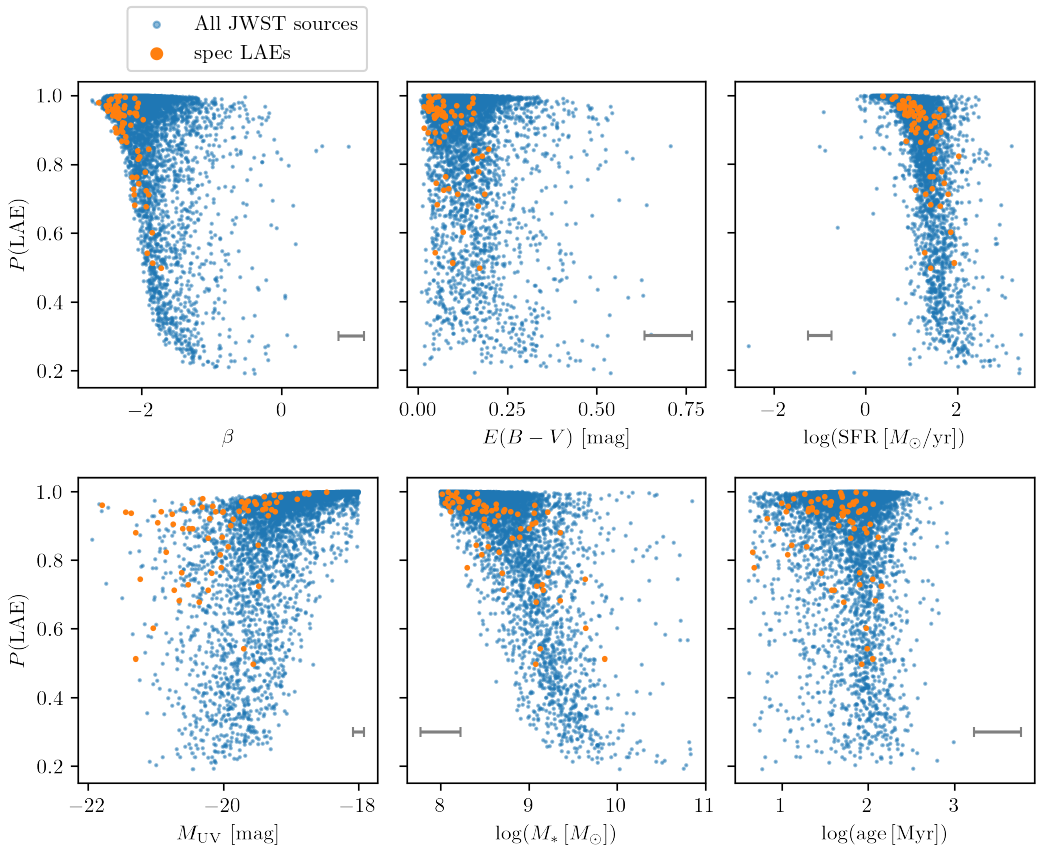}
    \caption{The relation between the six parameters and $P(\mathrm{LAE})$ inferred by the prediction model for all JWST galaxies (blue) and spectroscopically confirmed LAEs (orange) in the JWST fields.}\label{fig:jwst_param}
\end{figure*}

JWST has been exploring a deep and detailed view of distant galaxies in the middle of the reionization phase.
Narrow but deep optical--to--IR multi-color data is provided by JWST, but \lya{} optical spectroscopy at $z<7$ is limited, and \lya{} from the reionization epoch at $z>7$ is hampered by IGM absorption.
By applying our prediction model to this data, we not only confirm the presence or absence of \lya{} emission lines at $z<7$, but also predict whether galaxies in the reionization era originally emit \lya{} emission lines before undergoing IGM attenuation.
By observing \lya{} emission lines for the sample, a more accurate \lya{} fraction can be derived, which constrains the reionization history.
In principle, this method can determine whether the observed non-LAE was originally an LAE, thus it allows spatial mapping of cosmic neutral fraction \citep[e.g.,][]{Yoshioka22} to reveal the patchy reionization process.
We show the results of applying the predictive model to the post-reionization galaxies at $3 < z < 6$, which is the same redshift range of galaxies in the training dataset.
Then, we apply to reionization era galaxies and constrain the size of ionized bubbles at $z > 7$.

\subsubsection{LAE prediction}
We derive $P(\mathrm{LAE})$ of galaxies detected with JWST using the prediction model.
The correlation between $P(\mathrm{LAE})$ and the input parameters is shown as blue points in Fig.~\ref{fig:jwst_param}.
The galaxies tend to have higher $P(\mathrm{LAE})$ compared to the training dataset because JWST observations detect fainter galaxies.
Same as Sec.~\ref{sec:learning}, we define LAEs as galaxies predicted to have $P(\mathrm{LAE}) > 0.7$, and non-LAEs as those with $P(\mathrm{LAE}) < 0.7$.

\subsubsection{Comparison with Spectroscopically Confirmed LAEs}
There are several spectroscopically confirmed LAEs in JWST fields.
\citet{Stawinski23} present 126 LAEs at $2.8 < z < 6.5$ in the CEERS field.
\lya{} emission is detected using Keck/DEIMOS before the launch of JWST in 2021.
They are confirmed as LAEs by visual inspection.
After the advent of JWST, \citet{Jones23} and \citet{Saxena23} detect 16 LAEs at $5.6 < z \lesssim 8$ using JWST/NIRSpec observation.
\citet{Jones23} select LAEs with EW greater than 3$\sigma$ error.
We evaluate the prediction model using these spectroscopically confirmed LAE samples.
For \citet{Stawinski23}, 73 out of 126 LAEs are included in the JWST imaging catalogue, and the physical properties from the SED fitting described above are available for 64 galaxies with reduced $\chi^2<2.5$.
Four LAEs of \citet{Jones23} are included in the JWST imaging catalogue.
Comparing the spectroscopic redshift and photometric redshift of the LAEs, both redshifts agree well.
Predicted $P(\mathrm{LAE})$ of spectroscopically confirmed LAEs are also shown in Fig.~\ref{fig:jwst_param}.
62 (91\%) out of 68 spectroscopically confirmed LAEs have $P(\mathrm{LAE})$ higher than 0.7, satisfying the criteria defined above.
While this high success rate might come from the fact that most of the JWST LAEs are low-mass galaxies, which bias the prediction model towards LAEs, the high $P(\mathrm{LAE})$ values of galaxies with $M_* < 10^9\,M_\odot$ is reasonable given their blue UV slope $\beta < -2$.
When limiting LAEs with $M_* > 10^9\,M_\odot$, the prediction model shows 67\% success rate (12 out of 18).
In addition to the results in Sec.~\ref{sec:cosmos_sc4k}, this result shows our prediction model is applicable to galaxy samples without spectroscopy even with JWST observation.

\subsubsection{\lya{} fraction}
\begin{figure*}
    \centering
    \includegraphics[]{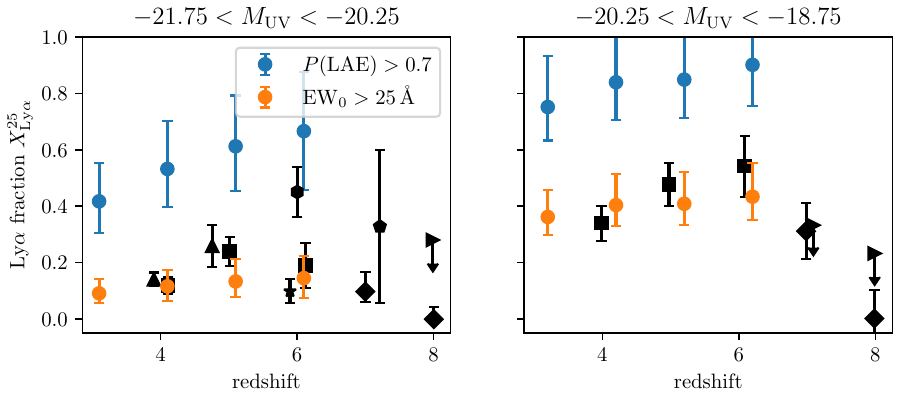}
    \caption{The redshift evolution of \lya{} fraction for $-21.75 < M_\mathrm{UV} < -20.25$ (left) and $-20.25 < M_\mathrm{UV} < -18.75$ (right).
    Blue points are calculated from LAEs selected with $P(\mathrm{LAE}) > 0.7$ from JWST galaxies.
    Orange points indicate the fraction of LAEs with $\mathrm{EW}_0 > 25 \,\mathrm{\AA}$ for JWST sources calibrated by the EW-$M_\mathrm{UV}$ relation of the training dataset.
    Black points show the previous results taken from \citet{Stark11, Schenker14, Ono12, Curtis-Lake12, Tilvi14, Cassata15}.}\label{fig:lya_fraction}
\end{figure*}

\lya{} fraction defined as a ratio of the number of LAEs to star-forming galaxies existing in the Universe is widely used to constrain the evolution of the cosmic neutral fraction in EoR.
Here we show the redshift evolution of \lya{} fraction at $3 < z < 6$ predicted by the model.
Because the EW distribution depends on $M_\mathrm{UV}$, previous studies draw the \lya{} fraction at $-21.75 < M_\mathrm{UV} < -20.25$ and $-20.25 < M_\mathrm{UV} < -18.75$ separately.
Many of the training dataset galaxies have similar $M_\mathrm{UV}$ to these two $M_\mathrm{UV}$ ranges, so we calculate \lya{} fraction of JWST galaxies with the prediction model in the same $M_\mathrm{UV}$ ranges.

Fig.~\ref{fig:lya_fraction} shows the evolution of \lya{} fraction.
We derive the uncertainty of \lya{} fraction arising from a Bernoulli trial and completeness and contamination in the LAE sample.
The uncertainty from the Bernoulli trial is given by a binomial proportion confidence interval \citep{Kusakabe20}.
The completeness and contamination are estimated by the performance of the prediction model (Sec.~\ref{sec:validation}).
\lya{} fraction is increasing at $3 < z < 6$ similar to the previous results while the amplitude is systematically higher.
This can be attributed to the difference in the definition of LAEs;
previous studies select LAEs that have EW$_0$ of \lya{} emission higher than $25 \,\mathrm{\AA}$, whereas we select LAEs by visual inspection.
In our selection, as described in Sec.~\ref{sec:training_data}, galaxies in the training dataset with smaller $\mathrm{EW}_0$ down to $3\,\mathrm{\AA}$ are also classified as LAEs.
This leads to the detection of more LAE by the prediction model, resulting in a higher \lya{} fraction in Fig.~\ref{fig:lya_fraction}.
To fairly compare the estimated \lya{} fraction with the previous studies, we estimate the percentage of LAEs with an $\mathrm{EW}_0$ greater than $25\,\mathrm{\AA}$ among all LAEs.
Using the training dataset consisting of VANDELS and MUSE galaxies, we calculate the number ratio of LAEs with $\mathrm{EW}_0 > 25 \,\mathrm{\AA}$ to total LAEs.
The number ratio of LAEs with $\mathrm{EW}_0 > 25 \,\mathrm{\AA}$ depends on the UV absolute magnitude of galaxies:
0.22 and 0.48 for $-21.75 < M_\mathrm{UV} < -20.25$ and $-20.25 < M_\mathrm{UV} < -18.75$, respectively.
We correct the \lya{} fraction with $\mathrm{EW}_0 > 25 \,\mathrm{\AA}$ of model predicted JWST LAEs by multiplying the above number ratios.
After the correction, the expected \lya{} fraction $X_\mathrm{Ly\alpha}^{25}$ are consistent with the previous results for both $-21.75 < M_\mathrm{UV} < -20.25$ and $-20.25 < M_\mathrm{UV} < -18.75$.
The \lya{} fraction reproduces the increasing trend at $z < 6$.
Unfortunately, the number of galaxies to which SED fitting can be applied is still too small to show model predictions of the \lya{} fraction without the IGM attenuation at $z>6$, but if this becomes possible in the future, it will help to understand the reionization history.

\subsubsection{LAE spatial distribution}
\begin{figure*}
    \centering
    \includegraphics[]{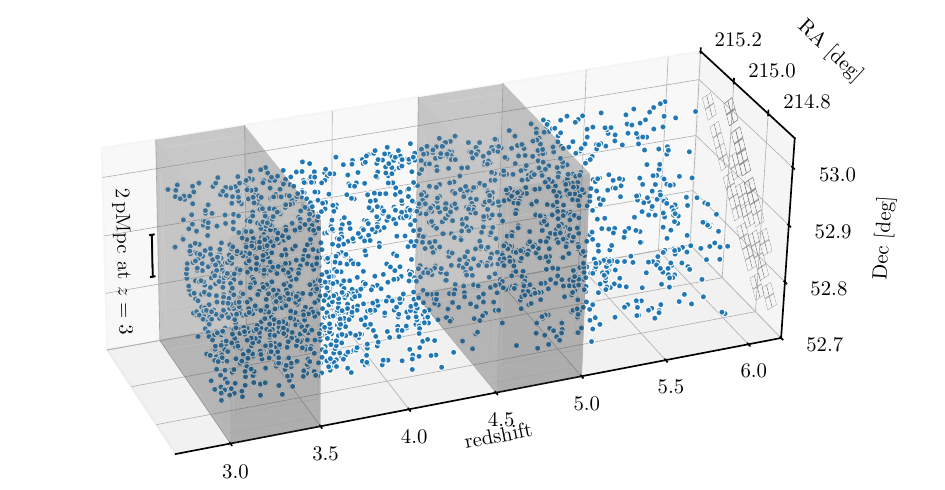}
    \includegraphics[]{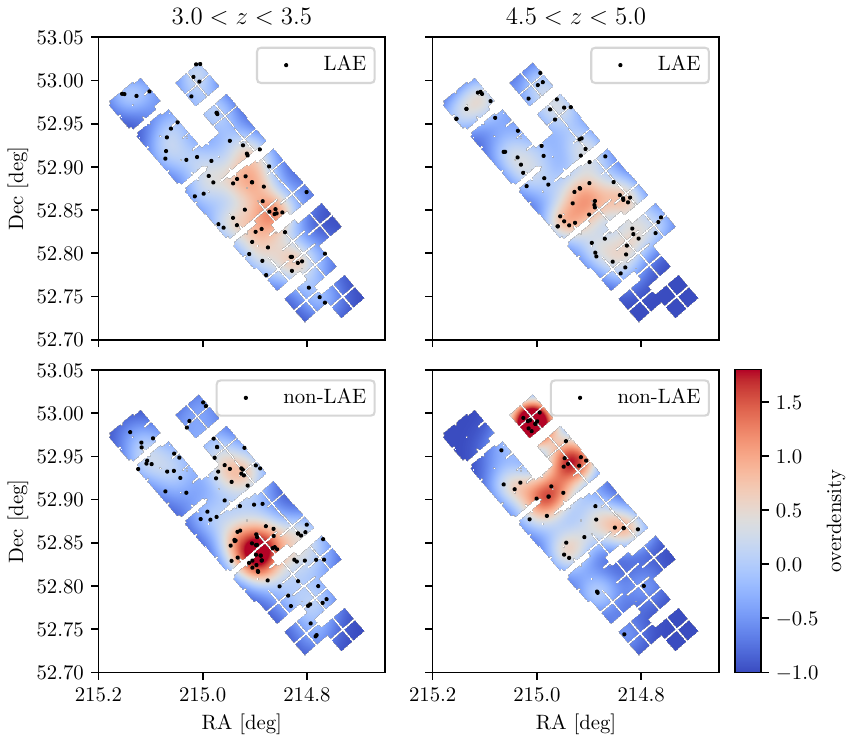}
    \caption{
        Top panel shows the 3D distribution of LAEs detected by the prediction model in the CEERS field.
        The grey shaded region indicates the redshift slices of $3.0 < z < 3.5$ and $4.5 < z < 5.0$, where the density distribution of LAEs and non-LAEs are shown in the bottom panels. 
        Bottom four panels show the density distribution of LAEs (upper) and non-LAEs (lower) in the CEERS field in the redshift slices of $3.0 < z < 3.5$ (left) and $4.5 < z < 5.0$ (right), respectively.
        The colour map represents the overdensity ($\delta = \sigma / \bar \sigma - 1$), where $\sigma$ is the density distribution calculated with the kernel density estimation. 
        The black points indicate LAEs (top) and non-LAEs (bottom).
    }\label{fig:lae_distribution}
\end{figure*}

Fig.~\ref{fig:lae_distribution} shows the 3D distribution of predicted LAEs in the CEERS field.
This demonstrates that the prediction model can detect LAEs in a wide redshift range continuously.
There seem to be gaps in the distribution of LAEs at $z=4.0$ and $5.5$.
This is because at $z=4.0$ and $5.5$ the number of galaxies included in the JWST catalog is small.
As described in Sec.~\ref{sec:validation}, the performance of the prediction model is independent of redshift.
The relation between the distribution of LAEs and underlying large-scale structure is thought to be the key in galaxy formation history.
Fig.~\ref{fig:lae_distribution} also shows the density distribution of LAEs and non-LAEs in redshift slices of $3.0 < z < 3.5$ and $4.5 < z < 5.0$.
As discussed in Sec.~\ref{sec:validation}, the network tends to predict $P(\mathrm{LAE}) > 0.8$ when galaxies are faint ($M_\mathrm{UV} > -19$).
To make a fair comparison of the distributions between LAEs and non-LAEs with the balanced numbers of galaxies, we only use galaxies with $M_\mathrm{UV} < -19.5$.
The density distribution is calculated using the kernel density estimation method with a Gaussian kernel.
The kernel size is determined by the typical clustering scale of LAEs.
The correlation length of clustering of LAEs at $z=3$ is $\sim 2\,\mathrm{cMpc}$ \citep{Ouchi10, Ito21}, which corresponds to $0.02\,\mathrm{deg}$ at $z=3$.
Interestingly, an LAE overdensity coincides with a non-LAE overdensity at $3.0 < z < 3.5$ while high-density regions with LAEs seem to be segregated from the non-LAE overdensities at $4.5 < z < 5.0$.
This result may provide an interesting example regarding the arguments over whether LAEs trace the underlying dark matter distribution.
\citet{Bielby16} show that the clustering properties of LAEs can be understood as those on the low-mass side of LBGs.
\citet{Sobral18} argue that bright LAEs are good tracers of the most overdensities in the Universe.
On the other hand,
\citet{Shimakawa17} report that LAEs are deficient in a protocluster core.
\citet{Momose21} propose a possible selection bias that \lya{} emissions behind H\,\textsc{i} IGM overdense regions  are hard to detect.
\citet{Ito21} find that the bias parameter of LAEs differs from those of general star-forming galaxies selected with photometric redshift.
It would be interesting to quantify the differences in the distribution of LAE and non-LAE, but that is beyond the scope of this paper.

\subsubsection{Constraints on the ionized bubble size at $z \sim 7.18,\ 7.49$ in the CEERS field}
\begin{figure*}
    \centering
    \includegraphics[]{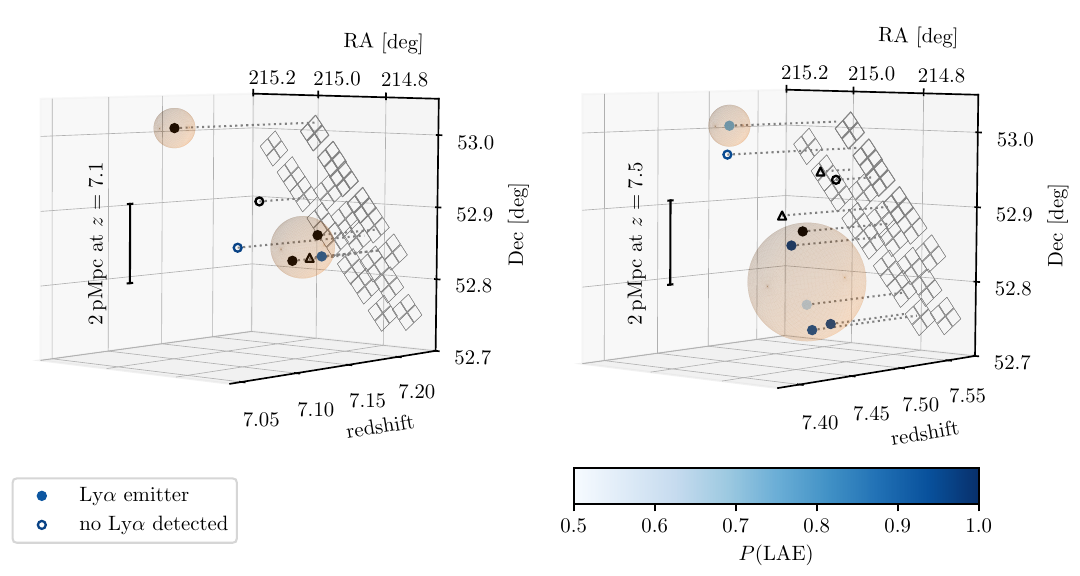}
    \caption{
        The distribution of spectroscopically confirmed galaxies around the expected ionized regions.
        The filled and open circles show galaxies with detected \lya{} emission and not detected in their spectra, respectively.
        The circles are coloured with the value of $P(\mathrm{LAE})$.
        The black circles show no $P(\mathrm{LAE})$ value because the galaxies are $M_* < 10^8 M_\odot$.
        The black open triangles show the galaxies with no UV counterpart in JWST/HST photometry or no spectral coverage at \lya{}.
        The orange shades are the expected ionized regions.
        The NIRCam footprint of the CEERS field is shown on the RA-DEC plane at the far end.
    }\label{fig:bubble}
\end{figure*}

In the reionization epoch, \lya{} emission is basically difficult to detect because it is attenuated by the neutral H\,\textsc{i} IGM.
Sometimes, however, \lya{} emission is observed even at $z>7$, especially in the overdense region of LAE, which is likely to be an ionized bubble.
\citet{Jung23}, \citet{Chen23}, and \citet{Napolitano24} detected \lya{} emission in the CEERS field at $z=7.1$ and $z=7.5$.
\citet{Chen23} expect large ionized bubbles ($R_\mathrm{ion} > 1\,\mathrm{pMpc}$),
which are thought to be carved out by the LAEs and surrounding fainter galaxies.
On the other hand, \citet{Napolitano24} suggest moderate size ionized regions ($R_\mathrm{ion} \lesssim 1\,\mathrm{pMpc}$) around LAEs.

In the moderately neutral Universe, \lya{} emission is attenuated by the IGM.
We assume the relation between galaxy physical properties and the \lya{} emission does not change with redshift, and our prediction model can predict whether galaxies in EoR intrinsically emit \lya{}.
Therefore, we can assess whether they have suffered IGM attenuation.
The comparison of the predicted LAE probability of each galaxy with the observed or unobserved \lya{} emission can constrain the size of the ionized bubble.
To achieve this goal, we use spectroscopically confirmed galaxies reported in \citet{Jung22}, \citet{Nakajima23}, \citet{Tang24}, \citet{Chen23}, and \citet{Napolitano24} in the CEERS field.
$P(\mathrm{LAE})$ of these galaxies are predicted using the network in the same way as Sec.~\ref{sec:JWST}, except for fixing redshift at spec-$z$ when the SED fitting.
We use only HST photometries for galaxies located outside of the NIRCam footprints.
We remove the galaxies with $M_* < 10^8M_\odot$ same as the training dataset.
Fig.~\ref{fig:bubble} shows the galaxy distribution with the predicted $P(\mathrm{LAE})$.

At $z = 7.18$, three LAEs are closely located, and one LAE is at $z = 7.10$.
The distance between the two structures is $4.3 \,\mathrm{pMpc}$ in the three-dimension space.
Two galaxies between these structures do not have \lya{} emission, even though one of them is predicted as LAEs with $P(\mathrm{LAE}) > 0.9$ and would have been intrinsically \lya{} emitting.
While its stellar mass is low ($10^{8.2}M_\odot$), it is reasonable that it is an intrinsic LAE given its UV slope $\beta$ ($-2.4$).
The lack of \lya{} emission in the observed spectra is attributed to the high neutral fraction in the IGM. 
Therefore, it is reasonable to consider that each of the LAE structures at $z = 7.10$ and $z = 7.18$ has a moderate-size ionized bubble ($R < 1 \,\mathrm{pMpc}$) rather than the galaxies within $z = 7.10\mbox{--}7.18$ reside in a single large ionized bubble.

Similarly at $z \sim 7.5$, six LAEs are reported across the CEERS field.
One LAE is located in the North-East, and five LAEs are within a distance of $2.2 \,\mathrm{pMpc}$ in the South-West.
Between the two structures, four spectroscopically confirmed galaxies exist, two of them do not show \lya{} emission in their spectra.
One galaxy without \lya{} emission is also intrinsically predicted as LAEs with $P(\mathrm{LAE}) > 0.9$.
Its stellar mass is $10^{8.8}M_\odot$, $M_\mathrm{UV}$ is $-20.7$, and UV slope $\beta$ is $-2.4$.
We regard the system at $z = 7.5$ consisting of at least two ionized regions while it is not clear whether the South-West structure is in a large ionized bubble or smaller ionized bubbles \citep{Napolitano24}.

\section{Summary}\label{sec:summary}
In this study, we develop a model to predict the probability that a galaxy shows \lya{} emission based on the neural network approach.
Our main results are summarized as follows.
\begin{enumerate}
    \item As a training dataset of the neural network, we collect spectroscopic information of \lya{} from VANDELS and MUSE spectroscopic surveys. We measure \lya{} flux and EW of \lya{} emission line. We conduct SED fitting using \texttt{CIGALE} to derive the physical properties of the galaxies. The spectra of galaxies are visually inspected to construct a training dataset of galaxies with and without \lya{} emission. In total, the training dataset consists of 1027 LAEs and 406 non-LAEs.
    \item We train a neural network that predicts whether a galaxy has a \lya{} emission line from six physical parameters, i.e., SFR, stellar mass, UV absolute magnitude $M_\mathrm{UV}$, age, UV slope $\beta$, and dust attenuation $E(B-V)$. We employ a Monte Carlo approach to account for the uncertainty in the input physical parameters. The trained prediction model shows the performance of 77\% true positive rate and 14\% false positive rate when we define LAEs as $P(\mathrm{LAE}) > 0.7$. The area under the ROC curve is 0.88.
    \item By the permutation feature importance method, we find that $\beta$, $M_\mathrm{UV}$, and $M_*$ have an impact on the prediction of LAEs.
    \item We apply the prediction model to COSMOS2020 sources and SC4K LAEs. Although these galaxies do not have spectroscopic information, our model outputs a reasonable $P(\mathrm{LAE})$ for each of them, thus validating its applicability to galaxy samples other than the training dataset.
    \item We use public JWST observations in the CEERS, COSMOS, GOODS-N, GOODS-S, and UDS fields to select LAEs with the prediction model. 91\% of the spectroscopically confirmed LAEs in the JWST fields are evaluated as $P(\mathrm{LAE}) > 0.7$, which indicates the validity of the prediction model. 
    \item We calculate \lya{} fraction of the JWST galaxies at $3 < z < 6$ based on the output $P(\mathrm{LAE})$ of the prediction model. The \lya{} fraction reproduces the increasing trend in this redshift range found in the previous studies. 
    \item Using LAEs detected by the prediction model from JWST observations, we show a continuous spatial distribution of LAEs over $3 < z < 6$ in the CEERS field. We find some similarities ($3.0 < z < 3.5$) and discrepancies ($4.5 < z < 5.0$) in the density distribution of LAEs and non-LAEs. We also investigate the galaxy distribution around the expected ionized bubbles at $z=7.18$ and $z=7.49$. The comparison between the predicted $P(\mathrm{LAE})$ and the spectroscopically observed \lya{} flux suggests that the ionized structures at $z=7.18$ and $z=7.49$ are comprised of separated moderate size ($R_\mathrm{ion} \lesssim 1 \,\mathrm{pMpc}$) ionized bubbles rather than surrounded by one large ($R_\mathrm{ion} > 2\,\mathrm{pMpc}$) ionized bubble. 
\end{enumerate}

\section*{Acknowledgements}
    We thank the anonymous referee for the constructive comments that improved the quality of the paper.
    This research was supported by IGPEES, WINGS Program, the University of Tokyo.
    This research was supported by a grant from the Hayakawa Satio Fund awarded by the Astronomical Society of Japan.
    N.K. was supported by the Japan Society for the Promotion of Science through Grant-in-Aid for Scientific Research 21H04490.
    Y.T. is supported by the Forefront Physics and Mathematics Program to Drive Transformation (FoPM), a World-leading Innovative Graduate Study (WINGS) Program, the University of Tokyo, Iwadare Scholarship Foundation and JSPS KAKENHI Grant Number JP23KJ0726.
    Some of the data products presented herein were retrieved from the Dawn JWST Archive (DJA). DJA is an initiative of the Cosmic Dawn Center (DAWN), which is funded by the Danish National Research Foundation under grant DNRF140.

\textit{Software:}
    adstex (\url{https://github.com/yymao/adstex}),
    Astropy \citep{astropy13, astropy18},
    Jupyter \citep{jupyter},
    Keras \citep{chollet2015keras},
    Matplotlib \citep{matplotlib},
    Numpy \citep{numpy},
    pandas \citep{pandas, McKinney10},
    Scikit-learn \citep{scikit-learn},
    SciPy \citep{scipy},
    Tensorflow \citep{tensorflow},
    uncertainties (\url{http://pythonhosted.org/uncertainties/})

\section*{Data Availability}
The data from VANDELS are available at \url{http://vandels.inaf.it/}.
The MUSE catalog of \citet{Schmidt21} is available at \url{https://cdsarc.u-strasbg.fr/viz-bin/cat/J/A+A/654/A80}.
The data from CANDELS are available at \url{https://archive.stsci.edu/hlsp/candels}.
The data from 3D-HST are available at \url{https://archive.stsci.edu/prepds/3d-hst/}.
The data from UVUDF are available at \url{https://archive.stsci.edu/hlsp/uvudf}.
The data from COSMOS are available at \url{https://cosmos.astro.caltech.edu/}.
The data from SC4K are available at \url{https://academic.oup.com/mnras/article/476/4/4725/4858393}.
The photometric catalogs of JWST are available at \url{https://dawn-cph.github.io/dja/index.html}.

\bibliography{lya_nn}{}
\bibliographystyle{mnras}

\bsp    
\label{lastpage}
\end{document}